\documentclass[11pt]{article}
\usepackage{times,fullpage,amsmath,amsthm,amssymb,hyperref,cleveref,graphicx,color,subcaption,subfiles,appendix}

\title{An Optimal Algorithm for\\ Higher-Order Voronoi Diagrams in the Plane:\\
The Usefulness of Nondeterminism}

\author{Timothy M. Chan\thanks{Department of Computer Science, University of Illinois at Urbana-Champaign, USA (tmc@illinois.edu). Work supported in part by NSF Grant CCF-2224271.} 
\and Pingan Cheng\thanks{Department of Computer Science, Aarhus University, Denmark (pingancheng@cs.au.dk)}
\and Da Wei Zheng\thanks{Department of Computer Science, University of Illinois at Urbana-Champaign, USA (dwzheng2@illinois.edu).}}
\date{}

\newtheorem{theorem}{Theorem}[section]
\newtheorem{lemma}[theorem]{Lemma}

\newtheorem{observation}[theorem]{Observation}

\newcommand{\eps}{\varepsilon}

\newcommand{\R}{\mathbb{R}}

\newcommand{\IGNORE}[1]{}

\newcommand{\lineseg}[1]{\overline{#1}}
\newcommand{\Cert}{\mathcal{X}}

\begin{document}

\maketitle
\begin{abstract}
We present the first optimal randomized algorithm for constructing the order-$k$ Voronoi diagram of $n$ points in two dimensions.  The expected running time is $O(n\log n + nk)$, which improves the previous, two-decades-old result of Ramos (SoCG'99) by a $2^{O(\log^*k)}$ factor.  To obtain our result, we (i)~use a recent decision-tree technique of Chan and Zheng (SODA'22) in combination with Ramos's cutting construction, to reduce the problem to \emph{verifying} an order-$k$ Voronoi diagram, and (ii)~solve the verification problem by a new divide-and-conquer algorithm using planar-graph separators.

We also describe a deterministic algorithm for constructing the $k$-level of $n$ lines in two dimensions in $O(n\log n + nk^{1/3})$ time, and constructing the $k$-level of $n$ planes in three dimensions in $O(n\log n + nk^{3/2})$ time.  These time bounds (ignoring the $n\log n$ term) match the current best upper bounds on the combinatorial complexity of the $k$-level.  Previously, the same time bound in two dimensions was obtained by Chan (1999) but with randomization.
\end{abstract}

\section{Introduction}

Given a set $P$ of $n$ points in $\R^2$,
the \emph{order-$k$ Voronoi diagram} is defined as the planar subdivision where
two points $q,q'\in \R^2$ belong to the same region iff $q$ and $q'$
have the same set of $k$ nearest neighbors in $P$ (each region of this subdivision is a convex polygon).
The problem of designing efficient algorithms to construct the
order-$k$ Voronoi diagram has a long history~\cite{Aurenhammer91,4m,EdelsbrunnerBOOK87,MulmuleyBOOK,PreparataS85}, and appeared in
Ian Shamos's original PhD thesis~\cite{Shamos78} that marked the beginning of computational geometry
(e.g., see unsolved problem 5
on page 206 in the thesis).
Surprisingly, the time complexity for this basic problem
has still not been fully resolved, even though optimal algorithms have long been known
(from the 70s, 80s, and 90s) for most of the other textbook
problems in two-dimensional computational geometry, including the convex hull, 
the standard (order-1) Voronoi diagram, line segment intersection, polygon triangulation, etc.~\cite{4m,PreparataS85}.

Table~\ref{tbl1} shows how extensively the problem has been studied in the past.
Shamos and Hoey (FOCS'75)~\cite{ShamosH75} were the first to define the
order-$k$ Voronoi diagram.  Lee~\cite{Lee82} gave the first algorithm, and also proved that the combinatorial complexity of the
diagram (i.e., the total number of vertices, edges, and regions) in $\R^2$ is $\Theta(nk)$ for all $k\le n/2$.
Agarwal, de Berg, Matou\v sek, and Schwarzkopf (SoCG'94)~\cite{AgarwalBMS98} gave
the first randomized algorithm that is within logarithmic factors from optimal:
the expected running time is $O(n\log^3 n + nk\log n)$.
Subsequently, Chan~\cite{Chan00} improved it to $O(n\log n + nk\log k)$; more generally,
by using shallow cuttings, he showed that any $T(n)$-time algorithm can be converted to an $O(n\log n + (n/k)T(k))$-time algorithm,
and so it suffices to focus on time bounds as a function of $n$ alone 
(Agarwal et al.'s algorithm achieved $T(n)=O(n^2\log n)$).
Finally, Ramos (SoCG'99)~\cite{Ramos99} modified Agarwal et al.'s randomized incremental
algorithm %as a cutting construction for $k$-levels and then 
and incorporated recursion (i.e., divide-and-conquer) to obtain an improved bound $T(n)=O(n^2 2^{O(\log^*n)})$,
where $\log^*n$ is the (slow-growing) iterated logarithm function; by combining 
with Chan's reduction, the expected running time in terms of $n$ and $k$ then became $O(n\log n + nk2^{O(\log^*k)})$.
Ramos's result has not been further improved since, and has remained the record for over two decades.

\paragraph{New result.}
The main result of the present paper is a new randomized algorithm that runs in
$O(n\log n + nk)$ expected time.  This result is tight for all $k\le n/2$, since
an $\Omega(n\log n)$ lower bound holds in any comparison-based model, and
an $\Omega(nk)$ lower bound trivially holds because the output size is $\Theta(nk)$, as mentioned
(and not just in the worst case, but always).  The same result (like many of the previous results) applies also to the \emph{farthest-point} order-$k$
 Voronoi diagram; we thus obtain optimal bounds for $k>n/2$ as well
by replacing $k$ with $n-k$, since the nearest-point order-$k$  Voronoi diagram is the 
same as the order-$(n-k)$ farthest-point Voronoi diagram.  Although some may regard an improvement of a
$2^{O(\log^*k)}$ factor as small, the result is important for providing
\emph{the first optimal solution} to a fundamental problem in classical computational geometry.

\begin{table}
\begin{tabular}{|ll|ll|}\hline
authors & & run time & \\\hline
Lee '82 & \cite{Lee82} & $O(nk^2\log n)$ & det.$^*$ \\
Edelsbrunner, O'Rourke, and Seidel (FOCS'83) %'86 
& \cite{EdelsbrunnerOS86} & $O(n^3)$ & det.$^*$\\
Edelsbrunner '86 & \cite{Edelsbrunner86} & $O(nk\sqrt{n}\log n)$ & det.\\
Chazelle and Edelsbrunner (SoCG'85) %'87 
& \cite{ChazelleE87} & $O(n^2 + nk\log^2n)$ & det.\\
Clarkson (STOC'86) %1987 
& \cite{Clarkson87} & $O(n^{1+\eps}k)$ & rand. \\
Aggarwal, Guibas, Saxe, and Shor (STOC'87) %'89
& \cite{AggarwalGSS89} & $O(n\log n+nk^2)$ 
& det.$^*$ \\
Aurenhammer and 
Schwarzkopf (SoCG'91) %1992 
& \cite{AurenhammerS92} & $O(nk\log^2n+nk^2)$ & rand.\ inc.\\
Mulmuley '91 & \cite{Mulmuley91} & $O(n\log n+nk^2)$ &  rand.$^*$\\
Boissonnat, Devillers, 
and Teillaud '93 & \cite{BoissonnatDT93} & $O(n\log n+nk^3)$ & rand.\ inc.$^*$\!\\
Agarwal, de Berg, Matou\v sek, and Schwarzkopf (SoCG'94)\!\! %1998 
& \cite{AgarwalBMS98} & $O(n\log^3n+nk\log n)$ & rand.\ inc.\\
Agarwal and Matou\v sek '95 & \cite{AgarwalM95} & $O(n^{1+\eps}k)$ & det.\\
Chan (FOCS'98) %2000 
& \cite{Chan00} & $O(n\log n+nk\log k)$ & rand.\\
%  &  & $O(nk\log^2 k\,(\log n/\log k)^{O(1)})$ & det.\\
Ramos (SoCG'99) & \cite{Ramos99} & $O(n\log n + nk2^{O(\log^*k)})$\!\!
& rand.\\
Chan and Tsakalidis (SoCG'15) & \cite{ChanT16} &  $O(n\log n+nk\log k)$ & det.\\
new & & $O(n\log n + nk)$ & rand.\\\hline
\end{tabular}
\caption{History of algorithms for the order-$k$ Voronoi diagram in $\R^2$
($k\le n/2$).  Note: ``det.'' is short for deterministic, ``rand.'' for randomized,
and ``rand.\ inc.'' for randomized incremental, and $^*$ indicates that
the algorithm computes all diagrams of order $1$ to $k$. (This table is adapted from \cite{Chan00}.)}\label{tbl1}
\end{table}

\paragraph{Technical challenges and overview.}
The techniques we use to obtain the result are also theoretically quite interesting, in our opinion.
The starting point is Chan and Zheng's work in SODA'22~\cite{ChanZ22}, which described
a decision-tree-based paradigm yielding
an $O(n^{4/3})$-time algorithm for \emph{Hopcroft's problem} in $\R^2$, improving a previous
algorithm by Matou\v sek~\cite{Matousek93} running in $n^{4/3}2^{O(\log^*n)}$ time.
It was observed that an efficient algebraic decision tree with $O(n^{4/3})$ height can be 
automatically converted to an algorithm with the same time bound: basically,
existing geometric divide-and-conquer techniques (\emph{cuttings}) allow us to reduce Hopcroft's
problem to subproblems of very small size, like $\log\log n$ or $\log\log\log n$, and for such a
small input size, we can afford to build the decision tree explicitly as preprocessing.
To obtain better decision tree upper bounds, Chan and Zheng formulated a ``Basic Search Lemma'',
which (very loosely speaking) states that searching among $r$ options actually can be done with \emph{constant}
amortized cost (instead of $O(r)$ or $O(\log r)$) in the algebraic decision tree model, in certain scenarios when $r$ is small and we face
multiple such search subproblems that ``originate'' from a common input set.
Put another way, for algebraic decision trees, we can support a mild form of \emph{nondeterminism}---we basically have the ability to guess which of the $r$ options is the answer, so long as we can  efficiently verify our guess.  The lemma may also be viewed as a generalization
of a known technique by Fredman~\cite{Fredman76} from the 70s on the decision-tree complexity of certain sorting problems.
Chan and Zheng applied the lemma to shave logarithmic
factors from the cost of
point location subproblems (geometric analog of binary searches), and thereby improve
the decision tree complexity of previous algorithms.

It is natural to try to apply the same paradigm to order-$k$ Voronoi
diagrams, to eliminate the similar-looking $2^{O(\log^*n)}$ factor from
Ramos's previous $n^2 2^{O(\log^*n)}$ time bound~\cite{Ramos99}.
However, the order-$k$ Voronoi diagram problem is very different from Hopcroft's.
Still, the problem can similarly
be self-reduced to subproblems of very small size, due to Ramos's divide-and-conquer scheme,
and thus an $O(n^2)$ decision tree bound would also translate to an $O(n^2)$ time bound.

But how do we design an $O(n^2)$-height decision tree for order-$k$ Voronoi diagrams?
The Basic Search Lemma doesn't seem to help speed up Agarwal, de Berg, Matou\v sek, and 
Schwarzkopf's algorithm with $T(n)=O(n^2\log n)$, since the $\log n$ factor there did not arise from
point location or binary search (but was due to technical reasons inherent to their probabilistic analysis).
And it doesn't seem to help shave all the logarithmic factors
from 
Chazelle and Edelsbrunner's earlier algorithm~\cite{ChazelleE87} with $T(n)=O(n^2\log^2n)$ either:
the $\log^2n$ factor there came up the use of Overmars and van Leeuwen's dynamic data structure
for planar convex hulls~\cite{OvermarsL81}, and the Basic Search Lemma could probably remove one log, but not both 
(we could alternatively save one log factor by using Brodal and Jacob's far-more-complicated, dynamic 2D convex hull structure~\cite{BrodalJ02}, but it is even less clear how the Basic Search Lemma could eliminate the remaining log there).

In our new solution, we will take the Search Lemma to the extreme: rather than guessing from among
a small number of options, we will guess the entire order-$k$ Voronoi diagram!
Although the number of possible diagrams is exponential, if we first make the problem size very small by
another application of Ramos's divide-and-conquer, then the number would still be acceptable.
This way, Chan and Zheng's paradigm allows us to reduce the original problem to the
\emph{verification problem}: given a point set and a diagram, decide whether it is the correct 
order-$k$ Voronoi diagram.
The existence of such a reduction to verification which doesn't increase the time bound is somewhat surprising.\footnote{For example, one could compare with the well known technique of \emph{parametric search}~\cite{Megiddo83}, which reduces an optimization problem to its corresponding decision problem, but involves searching for just one real value; here, we are searching for an entire order-$k$ Voronoi diagram!.
}

There have been some past works on verification or certification algorithms in the computational
geometry literature \cite{DevillersLPT98,McConnellMNS11,MehlhornNSSSSU99} (prompted by more practical concerns from algorithm engineering),
and simple algorithms have been designed for basic verification problems such as verifying convex polytopes, triangulations,
and the standard (order-1) Voronoi diagram.  However, verifying an order-$k$ Voronoi diagram
appears more difficult.  We describe a new algorithm that verifies the order-$k$ Voronoi diagram
in $O(n^2)$ time without extra logarithmic factors.  Our algorithm interestingly uses a
divide-and-conquer based on planar graph separators.  For classical problems in computational
geometry related to convex hulls
and Voronoi diagrams, it is far more common to see divide-and-conquer based on
cuttings, simplicial partitions, or Clarkson--Shor-style random sampling~\cite{CheongMR17,ClarksonS89,MulmuleyBOOK}; in contrast, divide-and-conquer 
algorithms
based on planar graph separators for such classical geometric problems are relatively rarer (but see \cite{AggarwalHL90,ChanC08,ChanT16,DehneDDFK97,Ramos01} for some examples).  We do not know how to apply separators
to directly construct the order-$k$ Voronoi diagram, but we are successful in using them to verify
a given diagram.

%verification algorithm is deterministic, no nondeterminism or randomness

Admittedly, the usage of Chan and Zheng's decision-tree paradigm is not likely to lead to practical algorithms,
but from the theoretical perspective, the entire solution is not long nor complicated.
We will keep the description mostly self-contained (without assuming knowledge of Chan and Zheng's
framework nor referring to the aforementioned Basic Search Lemma),
assuming only Ramos's divide-and-conquer,
planar separators, and some dynamic geometric data structures as black boxes.

\paragraph{More results.}
By a standard lifting transformation,
the construction of the order-$k$ Voronoi diagram of $n$ points in $\R^2$ is well known to be reducible to
the construction of the \emph{$k$-level} of $n$ planes in $\R^3$ that are tangent to
the paraboloid $z=-x^2-y^2$~\cite{4m, EdelsbrunnerS86}.
Our algorithm (like many of the previous algorithms)
can more generally compute the $k$-level of any set of $n$ planes
in $\R^3$ that are \emph{in convex position},
%tangent to an arbitrary convex surface, 
i.e., planes that all participate in the lower envelope.

For $n$ \emph{arbitrary} planes (not necessarily in convex position) in $\R^3$, determining the worst-case size
of the $k$-level is a well-known open problem in combinatorial geometry: the current best upper bound is $O(nk^{3/2})$
by Sharir, Smorodinsky, and Tardos~\cite{SharirST01}.
In Section~\ref{sec:det:klev}, we describe a deterministic algorithm that constructs the $k$-level of $n$ arbitrary planes in $\R^3$ in
$O(n\log n + nk^{3/2})$ time, which is thus the best worst-case bound attainable under
current knowledge on the combinatorial complexity of the $k$-level.
The best previous result has running time $O(n\log n + f\log^4k)$~\cite{AgarwalM95,Chan20},
where $f$ is the output size; although our new result is not output-sensitive, it avoids the
four logarithmic factors.  

When specialized to computing the $k$-levels of $n$ lines in $\R^2$, we also obtain a new 
deterministic algorithm that runs in $O(n\log n + nk^{1/3})$ time.  The current
best upper bound on the combinatorial complexity is $O(nk^{1/3})$ by Dey~\cite{Dey98}.
Previously, Chan~\cite{Chan99} gave a randomized algorithm achieving the same $O(n\log n + nk^{1/3})$
time bound (which improved Agarwal, de Berg, Matou\v sek, and Schwarzkopf's earlier
$O(n\log^2 n + nk^{1/3}\log^{2/3}n)$ randomized algorithm~\cite{AgarwalBMS98}), but derandomization 
of his algorithm seems difficult or impossible.

Our new deterministic $k$-level algorithms are obtained from a different approach, not relying
on decision trees but instead using a more traditional geometric divide-and-conquer based on hierarchical cuttings~\cite{Chazelle93}.

\paragraph{Applications.}
As higher-order Voronoi diagrams and $k$-levels are fundamental structures in computational geometry,
our new results have a number of applications.  We briefly mention two specific examples:

\begin{itemize}
\item Given a set $P$ of $n$ points in the plane and a number $k$, 
and we want to find a subset $Q\subset P$ of $k$ points
minimizing the \emph{variance} $\frac{1}{k}\sum_{q,q'\in Q}\|q-q'\|^2$.
Aggarwal et al.~\cite{AggarwalIKS91} showed that this problem can be reduced to
the construction of the order-$k$ Voronoi diagram, and so can now be solved
in $O(n\log n + nk)$ expected time.
\item Given $d$ point sets $P_1,\ldots,P_d$ of total size $n$ in $\R^d$, a \emph{ham-sandwich cut} 
is a hyperplane that has $\lfloor |P_i|/2\rfloor$ points of $P_i$ on either side.
Lo, Matou\v sek, and Steiger~\cite{LoMS94} gave an algorithm to construct a ham-sandwich cut in $\R^d$,
by using an algorithm for constructing $k$-levels of hyperplanes in $\R^{d-1}$ as a subroutine.
Consequently, our new results imply a deterministic $O(n^{4/3})$-time algorithm for ham-sandwich cuts in $\R^3$,
and a randomized $O(n^{5/2})$-time algorithm for ham-sandwich cuts in $\R^4$.
(For other applications of $k$-level construction, see also \cite{DurocherK09,GopalaM08}.)
\end{itemize}

\section{Preliminaries}
Let $P$ %=\{p_1, \dots, p_n\}$ 
be a set of $n$ points in $\R^d$.
The \emph{nearest-point (resp.\ farthest-point) order-$k$ Voronoi diagram} of $P$ is a partition of
the plane into regions, where two points are in the same region iff they have the same set of $k$ closest (resp.\ farthest) points in $P$.

Let $H$ be a set of $n$ hyperplanes in $\R^d$.
The \emph{level} of a point $q$ refers to the number of hyperplanes of $H$ strictly below $q$.
The \emph{$k$-level} of $H$ consists of all faces of the arrangement of $H$ that have level exactly $k$.

By a standard lifting transformation, 
computing the  nearest-point (resp.\ farthest-point) order-$k$ Voronoi diagram of a set of
$n$ points in $\R^2$ is equivalent to computing the
\emph{$k$-level} of a set 
%= \{h_1, \dots, h_n\}$ 
of $n$ 
planes in $\R^3$ tangent to the paraboloid $z=-x^2-y^2$ 
(resp.\ $z=x^2+y^2$); e.g., see \cite{4m,EdelsbrunnerS86}. 
Thus, in this paper, we will focus on computing the 
$k$-level for $n$ planes in $\R^3$ tangent to the  paraboloid, or more generally, for $n$ planes that are 
\emph{in convex (resp.\ concave) position}, i.e., for $n$ planes that all bound the lower (resp.\ upper) envelope.

It is known that the $k$-level of $n$ planes in convex position in $\R^3$ has combinatorial complexity $O(nk)$~\cite{ClarksonS89,Lee82}.
The $xy$-projection of the $k$-level of $n$ planes in $\R^3$ form a planar graph.  Thus, the $k$-level may be represented by a standard representation scheme for planar subdivisions (e.g., doubly connected edge lists)~\cite{4m,PreparataS85}, with $O(nk)$ pointers or $O(nk\log n)$ bits of space; each vertex of the $k$-level may be represented as a triple of pointers to its defining planes.

The following reduction by Chan~\cite{Chan00} shows that for $k$-level algorithms, it suffices to obtain good time bounds in terms of $n$ alone (i.e., it suffices to focus on the hardest case when $k=\Theta(n)$):

\begin{lemma}[\cite{Chan00}]\label{lem:k:sens} 
If there is a $T(n)$-time algorithm for computing the $k$-level of $n$ planes in general, convex, or concave position in $\R^3$, then
there is an $O(n\log n + (n/k)T(k))$-time algorithm for computing the $k$-level of $n$ planes in general, convex, or concave position respectively in $\R^3$.
%(n-k)-level?
\end{lemma}

Roughly, the above reduction follows from the use of \emph{shallow cuttings}~\cite{Mat92}: for any set of $n$ planes in $\R^3$, there exist a collection of $O(n/k)$ simplices covering all points of level at most $k$, such that each simplex intersects at most $n/k$ planes, and each simplex is unbounded from below.  To construct the $k$-level, we simply construct the $k$-level inside each simplex $\Delta$ of the cutting for the $O(n/k)$ planes intersecting~$\Delta$.  Efficient algorithms are known for finding a shallow cutting, taking $O(n\log n)$ expected time~\cite{Ramos99} or $O(n\log n)$ deterministic time~\cite{ChanT16}.

In the next two sections, we will describe an $O(n^2)$-time algorithm for the $k$-level of $n$ planes in convex position in $\R^3$ (the concave case can be addressed by negating the $z$ coordinates).  By the above reduction, an $O(n\log n + nk)$-time algorithm would then follow.

Our solution will consist of two parts: in Section~\ref{sec:reduce:to:verify} we reduce the problem of constructing the $k$-level to the problem of verifying the $k$-level, using nontrivial ideas based on decision trees, and in Section~\ref{sec:verify} we present an $O(n^2)$-time algorithm for the verification problem, using separators.

\section{Reduction to the  Verification Problem}\label{sec:reduce:to:verify}
In this section, we will present a reduction from the $k$-level problem 
to the problem of verifying the $k$-level.  To accomplish this, we build on ideas from Chan and Zheng's previous technique~\cite{ChanZ22} for Hopcroft's problem using decision trees, although these ideas will be streamlined and redescribed in a self-contained way.

We begin with the following lemma, implicitly obtained by Ramos \cite{Ramos99}, which gives a self-reduction of the $k$-level problem to smaller instances of logarithmic size:

\begin{lemma}[\cite{Ramos99}] \label{lem:ramosdivconq}
Computing the $k$-level of a set $H$ of $n$ planes in convex position in $\R^3$ self-reduces to
$O(n^2/\log^2n)$ instances of the problem for subsets of $H$ of $O(\log n)$ size, after spending
$O(n^2)$ expected time (using randomization).
\end{lemma}

Roughly, the lemma  follows by constructing a cutting into $O(n^2/\log^2n)$ simplices  covering the $k$-level such that each simplex intersects $O(\log n)$ planes: Ramos~\cite{Ramos99} obtained his construction by running a variant of Agarwal et al.'s randomized incremental algorithm~\cite{AgarwalBMS98}, but preemptively stopping the algorithm after $O(n/\log n)$ iterations to keep the expected running time bounded by $O(n^2)$.  Ramos then applied the lemma recursively to obtain his $O(n^2 2^{O(\log^*n)})$-time  randomized divide-and-conquer algorithm.

We first apply the lemma to reduce the problem to designing algorithms in the \emph{decision tree} setting.  
In the decision tree model, an algorithm may perform certain tests on the input.\footnote{In algebraic decision trees, the tests are evaluations of algebraic predicates over the input real numbers, but here we may use any test function with binary outcomes.}  All other operations that do not depend on the input have zero cost.  Different executions of the algorithm lead to different paths in the decision tree, where each node in the tree corresponds to a test.  The cost of the algorithm is the maximum total running time of the tests (or maximum total expected running time if the tests are done by randomized algorithms) over all paths of the tree.

\begin{lemma}\label{lem:reduce:to:decis:tree}
If there is an algorithm with $O(n^2)$ cost in the decision tree model for computing the $k$-level of $n$ planes in convex position in $\R^3$, and
the tree can be constructed in (say) doubly exponential time, then 
there is a randomized algorithm for computing the $k$-level of $n$ planes in convex position in $\R^3$ in $O(n^2)$ expected time.
\end{lemma}
\begin{proof}
By applying Lemma~\ref{lem:ramosdivconq} three times, we can
reduce the problem to $O(n^2/b^2)$ subproblems of size~$b$, with $b = O(\log \log \log n)$, in $O(n^2)$ expected time.
When the problem size $b$ is this small, 
%the time for constructing the decision tree is sublinear.
we can construct one decision tree for all problems of size $b$ in time sublinear in $n$.
Afterwards, each subproblem can be solved by following a path in that decision tree in $O(b^2)$ time.  The total time bound is
$O(n^2/b^2)\cdot O(b^2)=O(n^2)$.
\end{proof}

We now present our reduction of the $k$-level problem to the verification problem.
The input to the verification problem is the given set of $n$ planes, and a candidate $k$-level, which as mentioned can be represented using $O(n^2)$ pointers or $O(n^2\log n)$ bits of space.
%(see \Cref{sec:verify}).

\begin{theorem}\label{thm:reduce:to:verify} %\label{thm:decisiontree}
If there is an algorithm for verifying the $k$-level of $n$ planes in convex position in $\R^3$ in $O(n^2)$ time,
then there is a randomized algorithm for computing the $k$-level of $n$ planes in convex position in $\R^3$ in $O(n^2)$ expected time.
\end{theorem}
\begin{proof}
Let $(H,k)$ denote an instance of the problem of computing the $k$-level for a set $H$ of planes in convex position in $\R^3$.
By Lemma~\ref{lem:reduce:to:decis:tree},
%(or more accurately, a randomized variant of the lemma with expected running time), 
it suffices to describe an algorithm  
with $O(n^2)$ cost in the decision tree model on 
an instance $(H, k)$  where $|H| = n$.
%expected number of comparisons.

As each plane can be specified by three reals, we can view an input $H$ as a point $x_H$ in $\R^{3n}$.
Consider a comparison %$\gamma_{h_1,h_2,h_3,h}$ 
that 
tests if the point $h_1\cap h_2\cap h_3$
is above the plane $h$ for four given planes $h_1,h_2,h_3,h\in H$.
Let $\gamma_{h_1,h_2,h_3,h}$ be the set of all inputs for which this test is true;
this is a semialgebraic set in $\R^{3n}$ of constant degree.
Let $\Gamma$ denote the set of all these $O(n^4)$ semialgebraic sets.

\newcommand{\AAA}{\mathcal{A}}
We first build the entire
arrangement $\AAA(\Gamma)$ of $\Gamma$ in $\R^{3n}$
(this step does not involve looking at the actual input $H$ and so has zero cost in the decision tree model).
%(as doing so involves no  comparisons of the input $H$).
By the Milnor--Thom Theorem~\cite{Milnor64,Thom65}, $\AAA(\Gamma)$ 
has at most $|\Gamma|^{O(n)} = n^{O(n)}$ cells.
Throughout our %decision tree 
algorithm, 
we will perform operations that decrease the number of potential cells of $\AAA(\Gamma)$ that $x_H$ can be---we call these the \emph{active cells}.

By another twofold application of \Cref{lem:ramosdivconq}, we reduce the problem 
to smaller instances $(H_i, k_i)$ for $i=1,\dots, O(n^2/b^2)$ 
where each $|H_i| \le b$ for $b = O(\log\log n)$.

%We will describe the algorithm for solving $(H_i, k_i)$ given 

Suppose
that we have already solved the
subproblems $(H_j, k_j)$ 
for $j=1, \dots i-1$.
To solve the next subproblem $(H_i,k_i)$, we do the following:
\begin{enumerate}
\item Scan through the active cells and generate an answer for $(H_i,k_i)$ for each active cell.  (This step does not involve looking at the actual input and has zero cost in the decision tree model.  Note that inputs lying in the same cell of $\AAA(\Gamma)$ have the same $k_i$-level of $H_i$, since the level is determined by the outcomes of  comparisons of the type above.)
\item Pick an answer that is most popular among the answers from step~1.
\item Run the verification algorithm for this answer.  This test has $O(b^2)$ cost by assumption.
\item If the verification algorithm returns true, we have a correct answer for $(H_i,k_i)$.  
\item Otherwise, we compute an answer for $(H_i,k_i)$
by any polynomial-time algorithm with $b^{O(1)}$ %$O(b^2\log b)$ 
cost.
\end{enumerate}

An answer to each subproblem is represented as
$O(b^2)$ words or $O(b^2\log b)$ bits.
So, there are at most $B := 2^{O(b^2\log b)}$ 
many possible answers.
(Note that in step~1,
there may be multiple valid answers per cell, since the representation of a level need not be unique; we may pick an arbitrary valid answer per cell.)

If our guess in step~2 is correct, then we would have spent $O(b^2)$ time %comparisons
verifying our guess in each iteration.  % by \Cref{thm:verk-level}.  
The total cost for this part is
$O(b^2)\cdot O(n^2/b^2) = O(n^2)$.

On the other hand,
if our guess is wrong, then 
we know that 
the cells of $\AAA(\Gamma)$ that have answer equal to this guess do not contain $x_H$ and can be marked inactive, and so
we would have reduced the number of active
cells of $\AAA(\Gamma)$  by at least a factor of
$\frac{B-1}{B}$.
Thus, step~5 is done at most  
$O(\log_{B/(B-1)} |\AAA(\Gamma)|) = O(B\log |\AAA(\Gamma)|) = O(2^{O(b^2\log b)} n\log n)$ times in total over all iterations.
The total cost for this part is thus
$O(2^{O(b^2\log b)} n\log n) \cdot b^{O(1)}
 \le n^{1+o(1)}$, for $b=O(\log\log n)$.

So the total cost is $O(n^2)$.  To construct the decision tree, we try all $2^{O(n^2)}$ possible execution paths; at each node of the decision tree, step~1 naively takes $n^{O(n)}$ time.  The time needed to construct the $O(n)$-dimensional arrangement $\AAA(\Gamma)$ initially is $2^{n^{O(1)}}$~\cite{BasuPR96, BasuPR06}. Thus, the total construction time is at most $2^{n^{O(1)}} + 2^{O(n^2)}n^{O(n)}$, which is indeed sub-doubly-exponential.
\end{proof}

\section{Verification Algorithm}\label{sec:verify}
By Theorem~\ref{thm:reduce:to:verify}, it remains to describe an $O(n^2)$-time algorithm for verifying the $k$-level of $n$ planes in convex position in $\R^3$.  Simple algorithms have already been known for various basic verification problems, such as verifying standard (order-1) Voronoi diagrams, convexity of polytopes, and planarity of subdivisions \cite{DevillersLPT98,MehlhornNSSSSU99}: for such problems, it suffices to mainly check  for certain ``local'' conditions.  However, verification of the $k$-level appears more challenging, where local tests are insufficient.  We will present a divide-and-conquer algorithm using planar separators.

To simplify our exposition,
we define three auxiliary, almost vertical 
planes  
sufficiently far enough so that all vertices in the arrangement of $H$ are bounded in all non-vertical directions by these three planes.
For each of the three auxiliary planes, we create $n$ copies of the plane that are slightly perturbed and parallel to one another.  Let $H_b$ be these $3n$ planes, and let $H' = H\cup H_b$.
%which is still in convex position. 
These planes will be useful to ensure that the intersection of the $k$-level with any plane of $H$ is bounded (a property that we will need later in the proof of Observation \ref{lem:star_shaped}).
Initially, we can compute the 2D arrangement of $H$ within each of the three auxiliary planes in $O(n^2)$ time~\cite{4m}.
Afterwards, it is straightforward to generate the part of the $k$-level of $H$ bounded by all the planes in $H_b$, in $O(n^2)$ total time.
Thus, the given candidate $k$-level of $H$ can be extended to form a candidate $k$-level of $H'$.  We will describe how to verify this candidate $k$-level of $H'$.

The following property will be important, which crucially exploits the convex position assumption.

\begin{figure}
    \centering
    \includegraphics[width=0.49\textwidth,page=1]{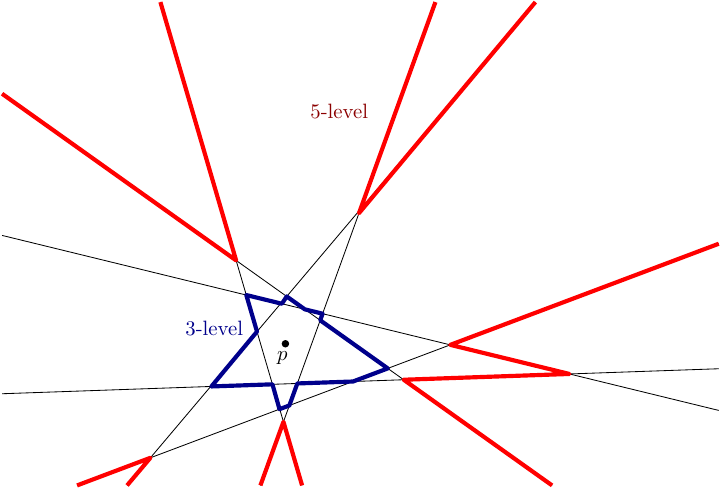}
    \includegraphics[width=0.49\textwidth,page=2]{star_modified.pdf}
    \caption{An example $3$-level and $5$-level intersecting one plane $h$ %with point $p_h$ on the $1$-level 
    before and after adding the bounding planes $H_b$. Observe that this plane originally lies below the $7$-level.}
    \label{fig:k-level}
\end{figure}

\begin{observation} \label{lem:star_shaped}
    Let $H$ be a set of $n$ planes in convex position in $\R^3$, and let $H'$ be defined above.  For any $h\in H$ and positive integer $k\le n$,
    the edges of the $k$-level of $H'$ involving the plane $h$ 
    %is either the empty set,  or 
    forms a cycle (in fact, a bounded star-shaped polygon).  Furthermore, we can find one point $p_h$ on this cycle, for every $h\in H$, in $O(n^2)$ total time.
\end{observation}
\begin{proof}
%We will restrict our attention to the single plane $h\in H$.
Let $p$ be a point on $h$ that appears on the lower envelope.
For some other point $q \in h$, 
suppose that the line segment $\lineseg{pq}$ between $p$ and $q$ 
intersect $k-1$ planes of $H$. Then $q$ lies on the $k$-level. 
Thus the set of points on the $k$-level on $h$ are exactly the set of points $q\in h$ such that $\lineseg{pq}$ intersect exactly $k-1$ lines. 
If one such point exists, it is possible to traverse the arrangement and find a cycle of points satisfying this condition, corresponding to a simple cycle of $G$, possibly using the planes of $H_b$.
We remark that our choice of $H_b$ guarantees at least one point exists, as
shooting a ray from $p$ in any direction will intersect at least $k$ planes.
\Cref{fig:k-level} illustrates this.
%one such point on the $k$-level always exists when $k\le |H|/2$ but it is possible that no such point exists when $k > |H|/2$.
%\DAVID{This is kind of messy, because these planes being vertical (or almost so) makes it very degenerate... meaning that for some $k$, the $k$-level is entirely on the bounding triangle. Furthermore, does this mean we have to add the degenerate part (which we can compute) to the graph?}

Finding one point on the cycle per $h$ in $O(n)$ time is straightforward, by computing intersections of $H$ with an arbitrary ray from $p$ on $h$.
%\TIMOTHY{In Figure 1, $p_h$ refers to a point in the kernel of the star-shaped polygon, but in the text, $p_h$ is a point on the polygon instead... all fixed now}
\end{proof}

Before we get into the details of our algorithm,
we first introduce two standard tools we will use.
First we need planar graph separators.  We will use the following version
by Klein, Mozes, and Sommer~\cite{KleinMS13}, which is obtained by recursively computing cycle separators to generate a decomposition tree:

\newcommand{\TTT}{\mathcal{T}}

\begin{lemma}[Decomposition tree from planar-graph  separators~\cite{KleinMS13}] \label{lem:separator}
Let $G$ be an embedded planar triangulated biconnected graph with $N$ vertices.
%and let $r$ be a constant.
In $O(N)$ time, we can build a rooted $O(1)$-degree tree $\TTT$, where each node $\nu$ is associated with a \emph{region} $R_\nu$ (a union of triangular faces and edges in $G$),\footnote{Many previous papers define regions edge-induced subgraphs, but we will think of them geometrically as sets in $\R^2$.} satisfying the following properties:\footnote{
Property (c) follows from equations (1) and (2) in Klein et al.'s paper~\cite{KleinMS13}, after compressing every 3 levels of their tree (the degree becomes 8).  Property (d) follows from Lemma~6 in \cite{KleinMS13} (which itself follows from a standard argument by Frederickson~\cite{Frederickson87}).
}
\begin{enumerate}
\item[(a)] the region of a node is the union of the regions of its  children, the regions of the  children are interior-disjoint, and the region at the root covers all triangular faces in $G$;
\item[(b)] each region is connected and has $O(1)$ holes;
\item[(c)] the complexity of a child's region is at most a constant fraction of the complexity of the parent's region;
\item[(d)] for any value $t$, the number of regions of complexity $\Theta(t)$ is $O(N/t)$, and the total complexity of their boundaries and their children's  boundaries is $O(N/\sqrt{t})$.
%nodes, and the region at each such node contains $O(r^i)$ vertices and has $O(r^{i/2})$ boundary edges.
\end{enumerate}
At each node $\nu$ of the tree, we store the boundary $\partial R_\nu$ of the region $R_\nu$ (and, for each boundary edge, a flag to indicate which side is ``inside'').
%    A \emph{decomposition tree} for an embedded planar graph $G$ is a rooted tree where
%    each node of the tree corresponds to a connected region of $G$ and stores a cycle separator.
%    If a node $v$ in the tree corresponds to a region with $n_v$ vertices, then the cycle seperator has size $O(\sqrt{n_v})$.
%    This tree can be constructed in $O(n)$ time.
\end{lemma}

Another tool we need is a data structure for 3D dynamic convex hulls, or in the dual, 3D dynamic lower/upper envelopes of planes.  The current best result by Chan \cite{Chan10,Chan20} stated below achieved $O(\log^4 n)$ amortized update time (although we can actually afford to use a weaker $O(n^\eps)$ update time bound~\cite{AgarwalM95}, or anything that is $o(n^{1/2-\eps})$):

%that supports vertical ray shooting queries.
\begin{lemma}[Ray shooting in 3D dynamic lower envelopes \cite{Chan10}]
\label{lem:dch}
There exists a data structure that handles insertions and deletions of up to $n$ planes in $\R^3$ in $O(\log^4n)$ amortized time, and given a query ray originating from inside the lower envelope, finds the point on the envelope hit by the ray  in $O(\log^2 n)$ time.
\end{lemma}

Now we are ready to present our verification algorithm.
\begin{theorem} \label{thm:verk-level}
 The $k$-level of a set $H$ of $n$ planes in convex position in $\R^3$ can be verified in $O(n^2)$ time.
\end{theorem}
\begin{proof}
Recall that the given candidate $k$-level of $H$ can be extended to a candidate $k$-level of $H'$, which we denote by $\Cert$.
Recall that $\Cert$ is stored using a standard representation scheme for planar subdivisions for its $xy$-projection. As a first step, we check that it is indeed a valid planar subdivision, i.e., the embedding does not have crossings.  We can just use a known verification algorithm by Mehlhorn et al.~\cite{MehlhornNSSSSU99} or Devillers et al.~\cite{DevillersLPT98} for this task, which takes time linear in the size of the subdivision (i.e., $O(n^2)$).
We assume that we have precomputed a point $p_h$
on $h$ that lies on the $k$-level of $H'$, for every $h\in H_\nu$, by 
\Cref{lem:star_shaped} in $O(n^2)$ total time.

Let $G_\Cert$ be a triangulation of the embedded planar graph formed by the $xy$-projection of $\Cert$.
The graph $G_\Cert$ has $N=O(n^2)$ size.
In $O(N)$ time, 
we compute the decomposition tree $\TTT$ for $G_\Cert$
by~\Cref{lem:separator}.

We will describe a recursive algorithm to verify that $\Cert$ is the $k$-level of $H'$ 
by using the decomposition tree~$\TTT$.
The input to the recursive algorithm is a node $\nu$ of $\TTT$, 
a subset $H_\nu\subseteq H$, and a number $k_\nu$,
where we are promised that the $k$-level of $H'$ coincides with the $k_\nu$-level of $H_\nu\cup H_b$ inside\footnote{``Inside'' here (and elsewhere in this proof) technically refers to the $xy$-projection.} the region $R_\nu$.
We want to verify that the portion of $\Cert$ inside the region $R_\nu$ coincides with the $k$-level of $H'$.
(Initially, at the root $\nu$, we take $H_\nu=H$ and $k_\nu=k$.)

\paragraph{Verifying the separator boundaries.}
Let $\{\nu_j\}_{j=1}^{O(1)}$ be the children of $\nu$.
We first verify that the boundary edges of each child region $R_{\nu_j}$, when lifted back to $\R^3$, are indeed on the $k$-level of $H'$, i.e., the $k_\nu$-level of $H_\nu\cup H_b$.
By property (b) in \Cref{lem:separator}, the boundary of $R_{\nu_j}$ has $O(1)$ components.
Take one component $\gamma$.
We pick an arbitrary vertex $v_s$ of $\gamma$ 
and compute its level in $H_\nu\cup H_b$ by iterating through all planes in $H_\nu$ in $O(|H_\nu|)$ time. 
If its level is not $k_\nu$, we reject.
Otherwise,
we follow a known approach used in previous output-sensitive $k$-level algorithms~\cite{EdelsbrunnerW86,Edelsbrunner86,AgarwalM95}:
We construct a dynamic ray shooting data structure
for the upper (resp.\ lower) envelope 
of the planes in $H_\nu$ below (resp.\ above) $v_s$ 
in time $O(|H_\nu|\log^4|H_\nu|)$ using Lemma~\ref{lem:dch}.  
To verify a neighboring vertex $v_g$ of $v_s$, 
we issue ray shooting queries along the edge connecting $v_s$ and $v_g$ 
to the sets of planes above and below $v_s$. 
This enables us to find the neighbor vertex of $v_s$ on the $k_\nu$-level. 
We verify if the neighbor vertex found is $v_g$, 
and if so we update the sets of planes to be those above and below $v_s$.
This includes the deletion of a plane and the insertion of a new plane.
We reject if we detect that any vertex of $\gamma$ is not on the $k_\nu$-level.
By Lemma~\ref{lem:dch}, each operation takes $O(\log^4|H_\nu|)$ amortized time (note that dynamic ray shooting for the planes in $H_b$ can be trivially done in $O(1)$ time without data structures).
So the entire process can be done in time $O((|H_\nu|+|\partial R_{\nu_j}|)\log^4 |H_\nu|)$.

\paragraph{Recursing in the child regions.}
Fix a child $\nu_j$.
We next verify that $\Cert$  is the $k$-level of $H'$ inside the child region $R_{\nu_j}$ by recursion.  To do so, we need to define $H_{\nu_j}$ and $k_{\nu_j}$.  To this end,
we classify each plane $h\in H_\nu$ as follows:
Call $h$ a \emph{boundary plane} if it defines a boundary edge of $R_{\nu_j}$.
Call $h$ an \emph{interior plane} if it is not a boundary plane and the point $p_h$ is inside $R_{\nu_j}$.
Call $h$ an \emph{exterior plane} if it is not a boundary plane and the point $p_h$ is outside $R_{\nu_j}$.
To determine which points $p_h$ are inside $R_{\nu_j}$, we can answer $|H_\nu|$  planar point location queries~\cite{4m,PreparataS85} in the region $R_{\nu_j}$ 
in total time $O((|H_\nu| + |\partial R_{\nu_j}|)\log |\partial R_{\nu_j}|)$.

By~\Cref{lem:star_shaped}, the intersection of any plane $h\in H_\nu$
with the $k$-level of $H'$ is a cycle and is thus connected.
Hence, exterior planes cannot participate in the $k$-level of $H'$ inside $R_{\nu_j}$ (as we have already verified the boundary edges of $R_{\nu_j}$).
We let $H_{\nu_j}$ contain all the  boundary planes and interior planes.  
We let $k_{\nu_j}$ be $k_\nu$ minus the number of exterior planes that are below an arbitrary point of $R_{\nu_j}$.
(Note that an exterior plane that is below an arbitrary point of $R_{\nu_j}$ will be below all points of $R_{\nu_j}$.)
Then we know that the $k$-level of $H'$ 
coincides with the $k_{\nu_j}$-level of $H_{\nu_j}$  inside $R_{\nu_j}$.
We can now  recursively solve the problem for the child $\nu_j$ with $H_{\nu_j}$ and $k_{\nu_j}$.

\paragraph{Running time analysis.}
Let $n_{\nu_j}$ be the number of interior planes as defined above for the node $\nu_j$.
We know that the number of boundary planes for the node $\nu_j$ is $O(|\partial R_{\nu_j}|)$.
Thus, $|H_{\nu}|\le n_\nu + O(|\partial R_\nu|)$ at all nodes $\nu$.
Let $b_\nu=|\partial R_\nu| + \sum_j |\partial R_{\nu_j}|$.
The cost at each node $\nu$ is bounded by
$O((n_\nu + b_\nu) \log^4 (n_\nu + b_\nu))$.

Let $\TTT_i$ be the nodes of $\TTT$ whose regions have size between $r^i$ and $r^{i+1}$ for a constant $r>1$.
By property (d) of Lemma~\ref{lem:separator}, $|\TTT_i|=O(N/r^i)$ and
$\sum_{\nu\in\TTT_i}b_\nu = O(N/r^{i/2})$.
Furthermore,
we know that $\sum_{\nu\in\TTT_i} n_\nu\le n$ (by disjointness of the regions in $\TTT_i$, since a node and its parent can't both in $\TTT_i$ by property~(c) if we pick $r>1$ sufficiently small).  
Trivially, $b_\nu \le O(|R_\nu|)=O(r^i)$  for all $\nu\in\TTT_i$.
%Recall that $r$ is a constant.
The total cost over all nodes in $\TTT_i$ is thus bounded by
\begin{align*}
    \sum_{\nu\in\TTT_i} (n_\nu+b_\nu)\log^4(n_\nu +b_\nu)\ 
    &\le\ O\left( n\log^4 N + \sum_{\nu\in\TTT_i}b_\nu\log^4 b_\nu + \sum_{\nu\in\TTT_i} b_\nu\log^4 n_\nu  \right) \\
    &\le\ O\left(n\log^4 N + \frac{N}{r^{i/2}} \log^4 (r^i) + \frac{N}{r^{i/2}} \log^4\left(\frac{\sum_{\nu\in\TTT_i} b_\nu n_\nu}{N/r^{i/2}}\right) \right) \\
        &\le\ O\left(n\log^4 N + \frac{i^4N}{r^{i/2}} + \frac{N}{r^{i/2}} \log^4\frac{r^i n}{N/r^{i/2}} \right) \\
    &\le O\left(n\log^4 N + \frac{i^4N}{r^{i/2}}\right),
\end{align*}
where 
the second inequality follows from Jensen's inequality,
and the last inequality follows from $N=\Omega(n)$.

The total running time is bounded by summing over all $i$:
\[
    \sum_{i=0}^{\log_r N} O\left(n\log^4N+\frac{i^4N}{r^{i/2}}\right)
    \ =\ O(n\log^5N + N)\ =\ O(n^2),
\]
since $N=O(n^2)$.
\end{proof}

Combining Theorem~\ref{thm:verk-level} with Theorem~\ref{thm:reduce:to:verify} and Lemma~\ref{lem:k:sens}, we conclude:

\begin{theorem}
The $k$-level of a set of $n$ planes in convex position in $\R^3$ can be computed in $O(n\log n + nk)$ expected time.  The same holds for the order-$k$ Voronoi diagram of a set of $n$ points in $\R^2$.
\end{theorem}

\section{Deterministic $k$-Level Algorithm}\label{sec:det:klev}
%\section{Deterministic $k$-Level Algorithm}

In this section, we describe a different approach to obtain deterministic algorithm for
constructing the $k$-level for an arbitrary set of lines in $\R^2$ or
planes in $\R^3$ (not necessarily in convex position).

Dey~\cite{Dey98} and Sharir, Smorodinsky, and Tardos~\cite{SharirST01}
proved that the combinatorial complexity of the $k$-level is upper-bounded
by $O(nk^{1/3})$ in $\R^2$ and $O(nk^{3/2})$ in $\R^3$ (the current best
lower bound is $n2^{\Omega(\sqrt{\log k})}$ in $\R^2$ and $nk2^{\Omega(\sqrt{\log k})}$ in $\R^3$, by T\'oth~\cite{Toth01}).  We will need a generalization of
these upper bounds for multiple consecutive levels, which are known and follow from the
same techniques (see~\cite{Dey98} in $\R^2$ and \cite{Chan10b} in $\R^3$):

\begin{lemma}\label{lem:klev}
Given $n$ lines in $\R^2$ and numbers $k$ and $j$,
the total combinatorial complexity of levels $k-j,\ldots,k+j$ is upper-bounded
by $O(n^{4/3}j^{2/3})$.

Given $n$ planes in $\R^3$ and numbers $k$ and $j$,
the total combinatorial complexity of levels $k-j,\ldots,k+j$ is upper-bounded
by $O(n^{5/2}j^{1/2})$.
\end{lemma}

The main tool we will use is a deterministic construction
of \emph{cuttings} that is sensitive to the number of vertices, due to Chazelle~\cite{Chazelle93}:

\begin{lemma}[Cutting lemma~\cite{Chazelle93}]\label{lem:cutting}
Let $H$ be a set of $n$ hyperplanes in $\R^d$ and
let $\Delta$ be a simplex.  Given~$r$, we can
cut $\Delta$ into $O(X_{\Delta} (r/n)^d + r^{d-1})$ interior-disjoint subsimplices, each
intersecting at most $n/r$ hyperplanes of $H$, where $X_{\Delta}$
denotes the number of vertices of the arrangement of $H$ that lie inside
$\Delta$.  The cutting can be constructed in $O(n)$ deterministic time if $r$ is a constant.
\end{lemma}

A hierarchy of cuttings can then be efficiently generated by applying
the above lemma recursively, as shown by Chazelle~\cite{Chazelle93}; such hierarchical
cuttings have led to many applications in range searching (e.g., \cite{ChanZ22,Matousek93}).
Here, we observe that this approach can lead to an efficient $k$-level
algorithm, just by a small variant where we recurse only in subsimplices relevant
to the $k$-level.  The resulting algorithm is simple to describe and analyze.
(It is a little surprising that this simple variant was overlooked in previous works on $k$-level algorithms.)

\paragraph{The algorithm.} 
Given a set $H$ of at most $n$ hyperplanes in $\R^d$, a number $k$, and a simplex $\Delta$,
we compute the $k$-level of $H$ inside $\Delta$ as follows (omitting trivial base cases):
\begin{enumerate}
\item Apply \Cref{lem:cutting} to cut $\Delta$ into subsimplices for 
a fixed constant $r$.
\item For each subsimplex $\xi$:
\begin{enumerate}
\item Let $H_{\xi}$ be the subset of at most $n/r$ hyperplanes of $H$ intersecting $\xi$.
\item Let $c_\xi$ be the number of hyperplanes of $H$ completely below $\xi$.
\item If $c_\xi \in [k-n/r,k]$, then recursively compute the $(k-c_\xi)$-level of $H_\xi$
inside $\xi$.
(If $c_\xi > k$ or $c_\xi<k-n/r$, then the level is empty inside $\xi$.)
\end{enumerate}
\item Combine the levels from all the subsimplices together.
\end{enumerate}

Steps 2(a) and 2(b) can be done naively in $O(n)$ time since $r$ is a constant.  Note that in step 2(c), if $c_\xi\in [k-n/r,k]$, all points in $\xi$ have level in $[c_\xi,c_\xi+n/r]
\subseteq [k-n/r,k+n/r]$ with respect to $H$.

The cost of Step~3 is at most $O(n^{d-1})$ (since the number of edges intersecting
the $(d-1)$-dimensional boundary of $\xi$ is at most $O((n/r)^{d-1})$).

\paragraph{Running time analysis.}

\newcommand{\CCC}{{\cal C}}

Let $N$ and $K$ denote the global value of $n$ and $k$ at the root of recursion.
At the $i$-th level of recursion, each subproblem has at most $N/r^i$
hyperplanes.
Let $\CCC_i$ be the collection of all simplices from the $i$-th level of recursion.
We know that the simplices $\CCC_i$ are disjoint and are contained
in levels $[K-N/r^i,K+N/r^i]$ (with respect to the global input set).
By Lemma~\ref{lem:klev}, the total number of vertices
in levels $[K-N/r^i,K+N/r^i]$ is bounded by $O(N^\alpha (N/r^i)^{d-\alpha})$,
where  $\alpha=4/3$ if $d=2$, and $\alpha=5/2$ if $d=3$.  Thus,
\begin{eqnarray*}
|\CCC_{i+1}| &=&  
O\left( \sum_{\Delta\in\CCC_i} \left(X_\Delta \left(\frac{r}{N/r^i}\right)^d \,+\, r^{d-1}\right) \right)\\
&\le& O\left( N^\alpha (N/r^i)^{d-\alpha}\cdot  \left(\frac{r}{N/r^i}\right)^d\right) \,+\, O(r^{d-1}) |\CCC_i|\\
&=& O( r^{i\alpha + O(1)})  + O(r^{d-1}) |\CCC_i|.
\end{eqnarray*}
Because $\alpha$ is strictly larger than $d-1$, 
the recurrence solves to $|\CCC_i|=O(r^{i\alpha})$, 
for a sufficiently large constant~$r$.

The total running time is
\[ O\left(\sum_{i=0}^{\log_r N} |\CCC_i|\cdot (N/r^i)^{d-1}\right)
\ \le\ O\left(\sum_{i=0}^{\log_r N} N^{d-1} r^{i(\alpha-(d-1))}\right)
\ =\ O(N^\alpha).
\]
We have thus obtained a deterministic algorithm running in $O(N^{4/3})$ time
for $d=2$, and $O(N^{5/2})$ time for $d=3$.
By the reduction in Lemma~\ref{lem:k:sens} (which holds in both $\R^2$ and $\R^3$ and is deterministic~\cite{ChanT16}),
%Chan's reduction~\cite{Chan00} (which can be derandomized
%using a deterministic algorithm for shallow cuttings~\cite{ChanT16}), 
we conclude:

\begin{theorem}
Given $n$ lines in $\R^2$ and a number $k$, there is a deterministic algorithm
that constructs the $k$-level in $O(n\log n + nk^{1/3})$ time.

Given $n$ planes in $\R^3$ and a number $k$, there is a deterministic algorithm
that constructs the $k$-level in $O(n\log n + nk^{3/2})$ time.
\end{theorem}

\section{Final Remarks}

%\paragraph{Remark.}
%\begin{itemize}
%\item
It is instructive to note why the hierarchical cutting approach in Section~\ref{sec:det:klev} does 
not work as well for the case of order-$k$ Voronoi diagrams in $\R^2$ or $k$-levels 
of planes in convex position in $\R^3$.  The reason is that the combinatorial complexity of the $k$-level here is quadratic, and so $\alpha=d-1=2$, which causes extra factors in the analysis.

%\item
In our optimal randomized algorithm for order-$k$ Voronoi diagrams in Sections \ref{sec:reduce:to:verify}--\ref{sec:verify}, %we note that 
the only place randomization is used is Ramos's divide-and-conquer (Lemma~\ref{lem:ramosdivconq}).  Our verification algorithm in Section \ref{sec:verify} is deterministic.

%\item
In our reduction to the verification problem (Theorem \ref{thm:reduce:to:verify}), it actually suffices to bound the cost of the verification algorithm in the decision tree model (not actual running time), and
we may even allow nondeterminism in the verification algorithm, even though we don't need to.  In other words, the certificate may contain more information besides the answer (the $k$-level) itself, so long as we can efficiently verify the certificate.
(With nondeterminism, some steps in the verification algorithm could be simplified; for example, we can avoid invoking a known algorithm to construct the planar-separator decomposition tree, by guessing all the separators, i.e., including them as part of the certificate; and point location also becomes easier with nondeterminism.)

%\item
The idea of reducing to verification, certification, or designing nondeterministic algorithms seems general and potentially applicable to other problems, although we don't have any other concrete applications at the moment.  Possible candidates include
Hopcroft's problem and affine degeneracy testing (given $n$ points in $\R^d$, decide whether there exist $d+1$ points lying on a common hyperplane): we could get faster algorithms for either problem if there exist efficient comparison-based algorithms to \emph{certify} no answers for Hopcroft's problem in $o(n^{4/3})$ time 
or affine degeneracy testing in $o(n^d)$ time (which we currently don't have).  

Note that the approach is applicable only for problems with superlinear complexity (due to an extra overhead cost of $n^{1+o(1)}$ in the proof of Theorem \ref{thm:reduce:to:verify}).  For example, we can't apply it to the minimum spanning tree (MST) problem despite the existence of linear-time MST verification algorithms.
%and we can't apply it to the largest empty rectangle problem to shave off the iterated logarithmic factor from the current best $O(n2^{O(\log^*n)}\log n)$ time bound \cite{??}.

%\end{itemize}

\small
\bibliographystyle{plainurl}
\bibliography{references}

\begin{thebibliography}{10}

\bibitem{AgarwalBMS98}
Pankaj~K. Agarwal, Mark de~Berg, Ji\v{r}\'{\i} Matou\v{s}ek, and Otfried
  Schwarzkopf.
\newblock Constructing levels in arrangements and higher order {V}oronoi
  diagrams.
\newblock {\em {SIAM} J. Comput.}, 27(3):654--667, 1998.
\newblock Preliminary version in SoCG 1994.
\newblock \href {https://doi.org/10.1137/S0097539795281840}
  {\path{doi:10.1137/S0097539795281840}}.

\bibitem{AgarwalM95}
Pankaj~K. Agarwal and Ji\v{r}\'{\i} Matou\v{s}ek.
\newblock Dynamic half-space range reporting and its applications.
\newblock {\em Algorithmica}, 13(4):325--345, 1995.
\newblock \href {https://doi.org/10.1007/BF01293483}
  {\path{doi:10.1007/BF01293483}}.

\bibitem{AggarwalGSS89}
Alok Aggarwal, Leonidas~J. Guibas, James~B. Saxe, and Peter~W. Shor.
\newblock A linear-time algorithm for computing the {V}oronoi diagram of a
  convex polygon.
\newblock {\em Discret. Comput. Geom.}, 4:591--604, 1989.
\newblock Preliminary version in STOC 1987.
\newblock \href {https://doi.org/10.1007/BF02187749}
  {\path{doi:10.1007/BF02187749}}.

\bibitem{AggarwalHL90}
Alok Aggarwal, Mark Hansen, and Frank~Thomson Leighton.
\newblock Solving query-retrieval problems by compacting {V}oronoi diagrams.
\newblock In {\em Proceedings of the 22nd Annual {ACM} Symposium on Theory of
  Computing (STOC)}, pages 331--340, 1990.
\newblock \href {https://doi.org/10.1145/100216.100260}
  {\path{doi:10.1145/100216.100260}}.

\bibitem{AggarwalIKS91}
Alok Aggarwal, Hiroshi Imai, Naoki Katoh, and Subhash Suri.
\newblock Finding {$k$} points with minimum diameter and related problems.
\newblock {\em J. Algorithms}, 12(1):38--56, 1991.
\newblock \href {https://doi.org/10.1016/0196-6774(91)90022-Q}
  {\path{doi:10.1016/0196-6774(91)90022-Q}}.

\bibitem{Aurenhammer91}
Franz Aurenhammer.
\newblock Voronoi diagrams - {A} survey of a fundamental geometric data
  structure.
\newblock {\em {ACM} Comput. Surv.}, 23(3):345--405, 1991.
\newblock \href {https://doi.org/10.1145/116873.116880}
  {\path{doi:10.1145/116873.116880}}.

\bibitem{AurenhammerS92}
Franz Aurenhammer and Otfried Schwarzkopf.
\newblock A simple on-line randomized incremental algorithm for computing
  higher order {V}oronoi diagrams.
\newblock {\em Int. J. Comput. Geom. Appl.}, 2(4):363--381, 1992.
\newblock Preliminary version in SoCG 1991.
\newblock \href {https://doi.org/10.1142/S0218195992000214}
  {\path{doi:10.1142/S0218195992000214}}.

\bibitem{BasuPR96}
Saugata Basu, Richard Pollack, and Marie{-}Fran{\c{c}}oise Roy.
\newblock On the combinatorial and algebraic complexity of quantifier
  elimination.
\newblock {\em J. {ACM}}, 43(6):1002--1045, 1996.
\newblock \href {https://doi.org/10.1145/235809.235813}
  {\path{doi:10.1145/235809.235813}}.

\bibitem{BasuPR06}
Saugata Basu, Richard Pollack, and Marie-Fran{\c{c}}oise Roy.
\newblock {\em Algorithms in Real Algebraic Geometry}, volume~10.
\newblock Springer, 2006.
\newblock \href {https://doi.org/10.1007/3-540-33099-2}
  {\path{doi:10.1007/3-540-33099-2}}.

\bibitem{BoissonnatDT93}
Jean{-}Daniel Boissonnat, Olivier Devillers, and Monique Teillaud.
\newblock A semidynamic construction of higher-order {V}oronoi diagrams and its
  randomized analysis.
\newblock {\em Algorithmica}, 9(4):329--356, 1993.
\newblock \href {https://doi.org/10.1007/BF01228508}
  {\path{doi:10.1007/BF01228508}}.

\bibitem{BrodalJ02}
Gerth~St{\o}lting Brodal and Riko Jacob.
\newblock Dynamic planar convex hull.
\newblock In {\em Proc. 43rd IEEE Symposium on Foundations of Computer Science
  (FOCS)}, pages 617--626, 2002.
\newblock \href {https://doi.org/10.1109/SFCS.2002.1181985}
  {\path{doi:10.1109/SFCS.2002.1181985}}.

\bibitem{Chan99}
Timothy~M. Chan.
\newblock Remarks on $k$-level algorithms in the plane.
\newblock 1999.
\newblock URL: \url{http://http://tmc.web.engr.illinois.edu/lev2d_7_7_99.ps}.

\bibitem{Chan00}
Timothy~M. Chan.
\newblock Random sampling, halfspace range reporting, and construction of
  {$(\le k)$}-levels in three dimensions.
\newblock {\em {SIAM} J. Comput.}, 30(2):561--575, 2000.
\newblock Preliminary version in FOCS 1998.
\newblock \href {https://doi.org/10.1137/S0097539798349188}
  {\path{doi:10.1137/S0097539798349188}}.

\bibitem{Chan10}
Timothy~M. Chan.
\newblock A dynamic data structure for 3-d convex hulls and 2-d nearest
  neighbor queries.
\newblock {\em J. {ACM}}, 57(3):16:1--16:15, 2010.
\newblock \href {https://doi.org/10.1145/1706591.1706596}
  {\path{doi:10.1145/1706591.1706596}}.

\bibitem{Chan10b}
Timothy~M. Chan.
\newblock On the bichromatic {$k$-set} problem.
\newblock {\em {ACM} Trans. Algorithms}, 6(4):62:1--62:20, 2010.
\newblock Preliminary version in SODA 2008.
\newblock \href {https://doi.org/10.1145/1824777.1824782}
  {\path{doi:10.1145/1824777.1824782}}.

\bibitem{Chan20}
Timothy~M. Chan.
\newblock Dynamic geometric data structures via shallow cuttings.
\newblock {\em Discret. Comput. Geom.}, 64(4):1235--1252, 2020.
\newblock Preliminary version in SoCG 2019.
\newblock \href {https://doi.org/10.1007/s00454-020-00229-5}
  {\path{doi:10.1007/s00454-020-00229-5}}.

\bibitem{ChanC08}
Timothy~M. Chan and Eric~Y. Chen.
\newblock In-place 2-d nearest neighbor search.
\newblock In {\em Proc. 19th Annual {ACM-SIAM} Symposium on Discrete Algorithms
  (SODA)}, pages 904--911, 2008.
\newblock URL: \url{http://dl.acm.org/citation.cfm?id=1347082.1347181}.

\bibitem{ChanT16}
Timothy~M. Chan and Konstantinos Tsakalidis.
\newblock Optimal deterministic algorithms for 2-d and 3-d shallow cuttings.
\newblock {\em Discret. Comput. Geom.}, 56(4):866--881, 2016.
\newblock Preliminary version in SoCG 2015.
\newblock \href {https://doi.org/10.1007/s00454-016-9784-4}
  {\path{doi:10.1007/s00454-016-9784-4}}.

\bibitem{ChanZ22}
Timothy~M. Chan and Da~Wei Zheng.
\newblock Hopcroft's problem, log-star shaving, {2D} fractional cascading, and
  decision trees.
\newblock In {\em Proc. 33rd {ACM-SIAM} Symposium on Discrete Algorithms
  (SODA)}, pages 190--210, 2022.
\newblock \href {https://doi.org/10.1137/1.9781611977073.10}
  {\path{doi:10.1137/1.9781611977073.10}}.

\bibitem{Chazelle93}
Bernard Chazelle.
\newblock Cutting hyperplanes for divide-and-conquer.
\newblock {\em Discret. Comput. Geom.}, 9:145--158, 1993.
\newblock Preliminary version in FOCS 1991.
\newblock \href {https://doi.org/10.1007/BF02189314}
  {\path{doi:10.1007/BF02189314}}.

\bibitem{ChazelleE87}
Bernard Chazelle and Herbert Edelsbrunner.
\newblock An improved algorithm for constructing {$k$th-}order {V}oronoi
  diagrams.
\newblock {\em {IEEE} Trans. Computers}, 36(11):1349--1354, 1987.
\newblock Preliminary version in SoCG 1985.
\newblock \href {https://doi.org/10.1109/TC.1987.5009474}
  {\path{doi:10.1109/TC.1987.5009474}}.

\bibitem{CheongMR17}
Otfried Cheong, Ketan Mulmuley, and Edgar Ramos.
\newblock Randomization and derandomization.
\newblock In Jacob~E. Goodman, Joseph O'Rourke, and Csaba~D. T{\'o}th, editors,
  {\em Handbook of Discrete and Computational Geometry}, pages 1159--1187.
  Chapman and Hall/CRC, 3rd edition, 2017.
\newblock URL: \url{http://www.csun.edu/~ctoth/Handbook/chap44.pdf}.

\bibitem{Clarkson87}
Kenneth~L. Clarkson.
\newblock New applications of random sampling in computational geometry.
\newblock {\em Discret. Comput. Geom.}, 2:195--222, 1987.
\newblock Preliminary version in STOC 1986.
\newblock \href {https://doi.org/10.1007/BF02187879}
  {\path{doi:10.1007/BF02187879}}.

\bibitem{ClarksonS89}
Kenneth~L. Clarkson and Peter~W. Shor.
\newblock Application of random sampling in computational geometry, {II}.
\newblock {\em Discret. Comput. Geom.}, 4:387--421, 1989.
\newblock Preliminary version in SoCG 1988.
\newblock \href {https://doi.org/10.1007/BF02187740}
  {\path{doi:10.1007/BF02187740}}.

\bibitem{4m}
Mark de~Berg, Otfried Cheong, Marc~J. van Kreveld, and Mark~H. Overmars.
\newblock {\em Computational Geometry: Algorithms and Applications}.
\newblock Springer, 3rd edition, 2008.
\newblock URL: \url{https://www.worldcat.org/oclc/227584184}.

\bibitem{DehneDDFK97}
Frank K. H.~A. Dehne, Xiaotie Deng, Patrick~W. Dymond, Andreas Fabri, and
  Ashfaq~A. Khokhar.
\newblock A randomized parallel three-dimensional convex hull algorithm for
  coarse-grained multicomputers.
\newblock {\em Theory Comput. Syst.}, 30(6):547--558, 1997.
\newblock Preliminary version in SPAA 1995.
\newblock \href {https://doi.org/10.1007/s002240000067}
  {\path{doi:10.1007/s002240000067}}.

\bibitem{DevillersLPT98}
Olivier Devillers, Giuseppe Liotta, Franco~P. Preparata, and Roberto Tamassia.
\newblock Checking the convexity of polytopes and the planarity of
  subdivisions.
\newblock {\em Comput. Geom.}, 11(3-4):187--208, 1998.
\newblock \href {https://doi.org/10.1016/S0925-7721(98)00039-X}
  {\path{doi:10.1016/S0925-7721(98)00039-X}}.

\bibitem{Dey98}
Tamal~K. Dey.
\newblock Improved bounds for planar {$k$-}sets and related problems.
\newblock {\em Discret. Comput. Geom.}, 19(3):373--382, 1998.
\newblock \href {https://doi.org/10.1007/PL00009354}
  {\path{doi:10.1007/PL00009354}}.

\bibitem{DurocherK09}
Stephane Durocher and David~G. Kirkpatrick.
\newblock The projection median of a set of points.
\newblock {\em Comput. Geom.}, 42(5):364--375, 2009.
\newblock \href {https://doi.org/10.1016/j.comgeo.2008.06.006}
  {\path{doi:10.1016/j.comgeo.2008.06.006}}.

\bibitem{Edelsbrunner86}
Herbert Edelsbrunner.
\newblock Edge-skeletons in arrangements with applications.
\newblock {\em Algorithmica}, 1(1):93--109, 1986.
\newblock \href {https://doi.org/10.1007/BF01840438}
  {\path{doi:10.1007/BF01840438}}.

\bibitem{EdelsbrunnerBOOK87}
Herbert Edelsbrunner.
\newblock {\em Algorithms in Combinatorial Geometry}.
\newblock Springer, 1987.
\newblock \href {https://doi.org/10.1007/978-3-642-61568-9}
  {\path{doi:10.1007/978-3-642-61568-9}}.

\bibitem{EdelsbrunnerOS86}
Herbert Edelsbrunner, Joseph O'Rourke, and Raimund Seidel.
\newblock Constructing arrangements of lines and hyperplanes with applications.
\newblock {\em {SIAM} J. Comput.}, 15(2):341--363, 1986.
\newblock Preliminary version in FOCS 1983.
\newblock \href {https://doi.org/10.1137/0215024} {\path{doi:10.1137/0215024}}.

\bibitem{EdelsbrunnerS86}
Herbert Edelsbrunner and Raimund Seidel.
\newblock Voronoi diagrams and arrangements.
\newblock {\em Discret. Comput. Geom.}, 1:25--44, 1986.
\newblock \href {https://doi.org/10.1007/BF02187681}
  {\path{doi:10.1007/BF02187681}}.

\bibitem{EdelsbrunnerW86}
Herbert Edelsbrunner and Emo Welzl.
\newblock Constructing belts in two-dimensional arrangements with applications.
\newblock {\em {SIAM} J. Comput.}, 15(1):271--284, 1986.
\newblock \href {https://doi.org/10.1137/0215019} {\path{doi:10.1137/0215019}}.

\bibitem{Frederickson87}
Greg~N. Frederickson.
\newblock Fast algorithms for shortest paths in planar graphs, with
  applications.
\newblock {\em {SIAM} J. Comput.}, 16(6):1004--1022, 1987.
\newblock \href {https://doi.org/10.1137/0216064} {\path{doi:10.1137/0216064}}.

\bibitem{Fredman76}
Michael~L. Fredman.
\newblock How good is the information theory bound in sorting?
\newblock {\em Theor. Comput. Sci.}, 1(4):355--361, 1976.
\newblock \href {https://doi.org/10.1016/0304-3975(76)90078-5}
  {\path{doi:10.1016/0304-3975(76)90078-5}}.

\bibitem{GopalaM08}
Harish Gopala and Pat Morin.
\newblock Algorithms for bivariate zonoid depth.
\newblock {\em Comput. Geom.}, 39(1):2--13, 2008.
\newblock \href {https://doi.org/10.1016/j.comgeo.2007.05.007}
  {\path{doi:10.1016/j.comgeo.2007.05.007}}.

\bibitem{KleinMS13}
Philip~N. Klein, Shay Mozes, and Christian Sommer.
\newblock Structured recursive separator decompositions for planar graphs in
  linear time.
\newblock In {\em Proc. 45th ACM Symposium on Theory of Computing (STOC)},
  pages 505--514, 2013.
\newblock \href {https://doi.org/10.1145/2488608.2488672}
  {\path{doi:10.1145/2488608.2488672}}.

\bibitem{Lee82}
Der{-}Tsai Lee.
\newblock On {$k$-}nearest neighbor {V}oronoi diagrams in the plane.
\newblock {\em {IEEE} Trans. Computers}, 31(6):478--487, 1982.
\newblock \href {https://doi.org/10.1109/TC.1982.1676031}
  {\path{doi:10.1109/TC.1982.1676031}}.

\bibitem{LoMS94}
Chi{-}Yuan Lo, Jir{\'{\i}} Matousek, and William~L. Steiger.
\newblock Algorithms for ham-sandwich cuts.
\newblock {\em Discret. Comput. Geom.}, 11:433--452, 1994.
\newblock \href {https://doi.org/10.1007/BF02574017}
  {\path{doi:10.1007/BF02574017}}.

\bibitem{Mat92}
Ji\v{r}\'{\i} Matou\v{s}ek.
\newblock Reporting points in halfspaces.
\newblock {\em Comput. Geom.}, 2:169--186, 1992.
\newblock \href {https://doi.org/10.1016/0925-7721(92)90006-E}
  {\path{doi:10.1016/0925-7721(92)90006-E}}.

\bibitem{Matousek93}
Ji\v{r}\'{\i} Matou\v{s}ek.
\newblock Range searching with efficient hierarchical cuttings.
\newblock {\em Discret. Comput. Geom.}, 10:157--182, 1993.
\newblock Preliminary version in SoCG 1992.
\newblock \href {https://doi.org/10.1007/BF02573972}
  {\path{doi:10.1007/BF02573972}}.

\bibitem{McConnellMNS11}
Ross~M. McConnell, Kurt Mehlhorn, Stefan N{\"{a}}her, and Pascal Schweitzer.
\newblock Certifying algorithms.
\newblock {\em Comput. Sci. Rev.}, 5(2):119--161, 2011.
\newblock \href {https://doi.org/10.1016/j.cosrev.2010.09.009}
  {\path{doi:10.1016/j.cosrev.2010.09.009}}.

\bibitem{Megiddo83}
Nimrod Megiddo.
\newblock Applying parallel computation algorithms in the design of serial
  algorithms.
\newblock {\em J. {ACM}}, 30(4):852--865, 1983.
\newblock \href {https://doi.org/10.1145/2157.322410}
  {\path{doi:10.1145/2157.322410}}.

\bibitem{MehlhornNSSSSU99}
Kurt Mehlhorn, Stefan N{\"{a}}her, Michael Seel, Raimund Seidel, Thomas Schilz,
  Stefan Schirra, and Christian Uhrig.
\newblock Checking geometric programs or verification of geometric structures.
\newblock {\em Comput. Geom.}, 12(1-2):85--103, 1999.
\newblock Preliminary version in SoCG 1996.
\newblock \href {https://doi.org/10.1016/S0925-7721(98)00036-4}
  {\path{doi:10.1016/S0925-7721(98)00036-4}}.

\bibitem{Milnor64}
John~W. Milnor.
\newblock On the {B}etti numbers of real algebraic varieties.
\newblock {\em Proc. Amer. Math. Soc.}, pages 275--280, 1964.

\bibitem{Mulmuley91}
Ketan Mulmuley.
\newblock On levels in arrangement and {V}oronoi diagrams.
\newblock {\em Discret. Comput. Geom.}, 6:307--338, 1991.
\newblock \href {https://doi.org/10.1007/BF02574692}
  {\path{doi:10.1007/BF02574692}}.

\bibitem{MulmuleyBOOK}
Ketan Mulmuley.
\newblock {\em Computational Geometry: An Introduction Through Randomized
  Algorithms}.
\newblock Prentice Hall, 1994.

\bibitem{OvermarsL81}
Mark~H. Overmars and Jan van Leeuwen.
\newblock Maintenance of configurations in the plane.
\newblock {\em J. Comput. Syst. Sci.}, 23(2):166--204, 1981.
\newblock \href {https://doi.org/10.1016/0022-0000(81)90012-X}
  {\path{doi:10.1016/0022-0000(81)90012-X}}.

\bibitem{PreparataS85}
Franco~P. Preparata and Michael~Ian Shamos.
\newblock {\em Computational Geometry: An Introduction}.
\newblock Springer, 1985.
\newblock \href {https://doi.org/10.1007/978-1-4612-1098-6}
  {\path{doi:10.1007/978-1-4612-1098-6}}.

\bibitem{Ramos99}
Edgar~A. Ramos.
\newblock On range reporting, ray shooting and {$k$}-level construction.
\newblock In {\em Proc. 15th Symposium on Computational Geometry (SoCG)}, pages
  390--399, 1999.
\newblock Long version at
  \url{https://citeseerx.ist.psu.edu/pdf/48510d7257565a167081e0629578ca10bd2c5296}.
\newblock \href {https://doi.org/10.1145/304893.304993}
  {\path{doi:10.1145/304893.304993}}.

\bibitem{Ramos01}
Edgar~A. Ramos.
\newblock An optimal deterministic algorithm for computing the diameter of a
  three-dimensional point set.
\newblock {\em Discret. Comput. Geom.}, 26(2):233--244, 2001.
\newblock Preliminary version in SoCG 2000.
\newblock \href {https://doi.org/10.1007/s00454-001-0029-8}
  {\path{doi:10.1007/s00454-001-0029-8}}.

\bibitem{Shamos78}
Michael~Ian Shamos.
\newblock {\em Computational Geometry}.
\newblock PhD thesis, Yale University, 1978.
\newblock URL:
  \url{http://euro.ecom.cmu.edu/people/faculty/mshamos/1978ShamosThesis.pdf}.

\bibitem{ShamosH75}
Michael~Ian Shamos and Dan Hoey.
\newblock Closest-point problems.
\newblock In {\em Proc. 16th IEEE Symposium on Foundations of Computer Science
  (FOCS)}, pages 151--162. {IEEE} Computer Society, 1975.
\newblock \href {https://doi.org/10.1109/SFCS.1975.8}
  {\path{doi:10.1109/SFCS.1975.8}}.

\bibitem{SharirST01}
Micha Sharir, Shakhar Smorodinsky, and G{\'{a}}bor Tardos.
\newblock An improved bound for \emph{k}-sets in three dimensions.
\newblock {\em Discret. Comput. Geom.}, 26(2):195--204, 2001.
\newblock \href {https://doi.org/10.1007/s00454-001-0005-3}
  {\path{doi:10.1007/s00454-001-0005-3}}.

\bibitem{Thom65}
Ren\'e Thom.
\newblock Sur l'homologie des vari\'et\'es alg\'ebriques re\'elles.
\newblock In S.~S. Cairns, editor, {\em Differential and Combinatorial
  Topology}. Princeton Univ. Press, 1965.

\bibitem{Toth01}
G{\'{e}}za T{\'{o}}th.
\newblock Point sets with many {$k$-}sets.
\newblock {\em Discret. Comput. Geom.}, 26(2):187--194, 2001.
\newblock \href {https://doi.org/10.1007/s004540010022}
  {\path{doi:10.1007/s004540010022}}.

\end{thebibliography}

%\section{Appendices}
%\subfile{sections/appendices}

\end{document}